\documentstyle[preprint,aps]{revtex}
\tightenlines

\newcommand{\be}{\begin{equation}}
\newcommand{\ee}{\end{equation}}
\newcommand{\ba}{\begin{eqnarray}}
\newcommand{\ea}{\end{eqnarray}}
\begin{document}
\draft

\title{Moving quantum agents in a finite environment} 

\author{Ilki~Kim and G\"{u}nter~Mahler}

\address{Institut f\"{u}r Theoretische Physik, Universit\"{a}t Stuttgart\\ 
           Pfaffenwaldring 57, 70550 Stuttgart, Germany}

\date{\today}

\maketitle

\begin{abstract}
 We investigate an all-quantum-mechanical spin network, in which a subset of 
 spins, the $K$ ``moving agents'', are subject to local and pair unitary 
 transformations controlled by their position with respect to a fixed ring of 
 $M$ ``environmental''-spins. We demonstrate that a ``flow of coherence'' 
 results between the various subsystems. Despite entanglement between the 
 agents and between agent and environment, local (non-linear) invariants may 
 persist, which then show up as fascinating patterns in each agent's 
 Bloch-sphere. Such patterns disappear, though, if the agents are controlled 
 by different rules. Geometric aspects thus help to understand the interplay 
 between entanglement and decoherence.
\end{abstract}


\narrowtext

\section*{Introduction}
Originally, coherence refers to a strictly defined phase-relationship between 
(linear) wave-components. Such phase-relationships also underly superposition 
states in quantum mechanics (with respect to a given set of basis states). 
The superposition principle can be extended to 
composite systems ($N$ subsystems), for which 
we usually refer to product-states as the pertinent basis set. 
Corresponding coherent superpositions then include ``local coherent 
states'' in which case the product character remains untouched, and 
so-called ``entangled'' states~\cite{SCH95} for which the strict product form 
is lost: The reduced subsystem-states appear ``non-pure'' for the latter 
case, entanglement acts as a source of local entropy. As opposed to local 
coherence, entanglement has no classical analogue. 

Decoherence may generally be seen as a suppression of superposition. In the 
strong sense this suppression results from (approximate) disjointness due to 
super-selection rules (leading to a classical domain proper~\cite{PRI98}). 
The effect of superposition may (partially) be quenched for collective 
observables (such as the total spin), when, due to phase-randomness, 
the individual 
spins, though each well-defined, virtually cancel each other (in this case 
coherence may be recovered, {\it{cf.}} spin 
echo~\cite{RHI71}). Roughly speaking, 
non-selective local properties of such incoherent ensembles are 
``ill-defined'', when used as an input to appropriate interferometers, no 
interference fringes are observed. ``Ill-defined'' properties also occur in 
individual quantum objects as a result of incompatibility 
({\it{cf.}} Heisenberg uncertainty relations). Entanglements in composite 
systems implies that 
local properties become ``incompatible'' for the state considered, the 
superposition character of reduced subsystem-states is 
suppressed~\cite{SCH95}: Local 
coherence may be said ``to have moved where nobody looks'' (while coherence 
as such has not been destroyed). In this paper we investigate local state 
changes resulting from an iterative map applied to a composite system whose 
state evolves in the (in terms of $N$) exponentially large 
Hilbert-space~\cite{FEY82}. This map may be visualized as arising from a 
circular motion of spins (``agents'') 
on a closed chain of fixed environment spins.

\section*{Definition of network space}
The quantum network~\cite{MAH98} to be considered here is composed of $N$ 
pseudo-spins with state $|j\rangle\!^{(\mu)};\,j=0,1;\,\mu=1,2, \cdots, N$, 
and the transition-operators $\hat{P}_{ij}^{(\mu)}=|i\rangle\!^{(\mu)} 
{}^{(\mu)}\hspace*{-0.8mm}
\langle j|$. 
The $2^{N}$ product-states are 
\be 
|jk \cdots l\rangle = |j\rangle\!^{(1)} \otimes |k\rangle\!^{(2)} \otimes 
\cdots \otimes |l\rangle\!^{(N)}
\ee
on which the ``$c$-cluster-operators'' 
\be
\hat{Q}_{jk \cdots l} := \hat{\lambda}_{j}^{(1)} \otimes 
\hat{\lambda}_{k}^{(2)} \otimes \cdots \otimes \hat{\lambda}_{l}^{(N)}
\ee
act. Here and in the following the upper indices (in parenthesis) refer to 
the subsystem-numbers, lower indices to the component/type of property. 
The $\hat{\lambda}_{i}^{(\mu)}$ are the following $SU(2)$-operators
\ba
\hat{\lambda}_{1}^{(\mu)} &=& \hat{P}_{01}^{(\mu)}+\hat{P}_{10}^{(\mu)}
\nonumber\\
\hat{\lambda}_{2}^{(\mu)} &=& i\bigl(\hat{P}_{01}^{(\mu)}-\hat{P}_{10}^{(\mu)}
\bigr)\nonumber\\
\hat{\lambda}_{3}^{(\mu)} &=& \hat{P}_{11}^{(\mu)}-\hat{P}_{00}^{(\mu)}
\nonumber\\
\hat{\lambda}_{0}^{(\mu)} &=& \hat{1}^{(\mu)} = \hat{P}_{11}^{(\mu)}+
\hat{P}_{00}^{(\mu)}
\ea
and $c$ is the number of indices $\ne 0$ within the given set of the lower 
indices $\{j k \cdots l\}$. $\,\hat{Q}_{jk \cdots l}$ thus operates on 
$\,c\,$~subsystems out of $\,N$. 
Full description of the network-state~$|\psi\rangle$ requires the 
specification of all the $\,{2^{2N}-1}$~expectation values
\be
-1 \leq \langle\psi|\hat{Q}_{jk \cdots l}|\psi\rangle =: Q_{jk \cdots l} 
\leq 1\;.
\ee
Here, $Q_{00 \cdots 0}=1$, the $\,c=1$~expectation values are the so-called 
Bloch-vectors for the individual spins, the $c > 1$ terms constitute 
$c$-point correlation functions. Dispersion-free expectation values are 
$\pm 1$.\,Largest quantum mechanical uncertainty pertains to operators with 
expectation-value zero. 
Therefore, as a convenient measure for the ``weight'' of well-defined 
properties of a given cluster of type~$c\,$ we take the respective 
``cluster-sum''~\cite{MAH98}
\ba
&c=1&:\;Y_{1}^{(1)}=\sum_{i=1}^{3}Q_{i00 \cdots 0}\,Q_{i00 \cdots 0},\;\;\;
        Y_{1}^{(2)}=\sum_{i=1}^{3}Q_{0i0 \cdots 0}\,Q_{0i0 \cdots 0}\;\;\;
        \mbox{etc.}\nonumber\\
&c=2&:\;Y_{2}^{(12)}=\sum_{i,j=1}^{3}Q_{ij0 \cdots 0}\,Q_{ij0 \cdots 0}\;\;\;
        \mbox{etc.}\nonumber
\ea
The $Y_{1}$-terms are identical with the square of the respective 
Bloch-vector-length. These cluster-sums obey the inequalities 
\be
Y_{c} \leq Z_{c} = \left\{ 
 \begin{array}{ll}
 2^{(c-1)} &\;\;\;\; c=\mbox{odd}\\ 
 2^{(c-1)}+2 &\;\;\;\; c=\mbox{even}\;.
\end{array}\right.
\ee
For a product-state each cluster-sum equals $1$, irrespective of its size. 
There are $2^{N}-1$~different clusters, implying
\be
2^{N}-1 = \sum_{c=1}^{N}\sum_{\mu,\nu \cdots} Y_{c}^{(\mu, \nu \cdots)}\;.
\ee
One easily shows that this sum-rule must hold for {\em{any}} pure state! 
As a consequence only few clusters can exploit their full allowance of $Z_{c}$ 
(which increases exponentially with $c$~!); in most cases $Y_{c} < Z_{c}$, 
which we call (partial) ``$c$-decoherence''. A given $2$-cluster has surplus 
correlation (``entanglement''), if for its partition 
$Y_{2}^{(12)} > Y_{1}^{(1)}Y_{2}^{(1)}$ holds. The various clusters have thus 
to ``compete for weight''. Dynamically we may say that coherence ``flows'' 
from one part to another (without getting lost, though), under the influence 
of unitary transformations.

\section*{Architecture}
We first split our network into $2$ subgroups, $N=K+M$, where $M$ pseudo-spins 
constitute the ``environment'', $\mu=1,2, \cdots, M,$ and the remaining spins 
are the ``agents'', $Sk=S1, S2,  \cdots, SK$ (Fig~\ref{architecture}). 
The behavior of the network is 
specified in terms of a discrete set of unitary 
transformations~\cite{MK98,KIM99},
\ba
\hat{U}_{\alpha_{k}}^{(Sk)} &=& \hat{1}^{(Sk)} \cos{(\alpha_{k}/2)} - 
\hat{\lambda}_{1}^{(Sk)} i \sin{(\alpha_{k}/2)}\nonumber\\
\hat{U}_{0}^{(Sk,\mu)} &=& \hat{P}_{00}^{(Sk)} \otimes 
\hat{\lambda}_{1}^{(\mu)} + \hat{P}_{11}^{(Sk)} \otimes \hat{1}^{(\mu)}
\nonumber\\
\hat{U}_{\pi}^{(Sk,\mu)} &=& \hat{P}_{00}^{(Sk)} \otimes 
i \hat{\lambda}_{2}^{(\mu)} + \hat{P}_{11}^{(Sk)} \otimes \hat{1}^{(\mu)}\;.
\ea
$\hat{U}_{\alpha_{k}}^{(Sk)}$ is a local rotation around the $x$~axis by 
angle $\alpha_{k}$, 
$\hat{U}_{\theta}^{(Sk,\mu)}$ is the quantum-controlled-NOT-operation (QCNOT) 
with 
$(\hat{U}_{\theta})^{2} = \hat{1}$ for $\theta=0$,\,
$(\hat{U}_{\theta})^{4} = \hat{1}$ for $\theta=\pi$. 
The sequence of transformations may be 
interpreted to result from cyclic and discretized~\cite{LLO96} 
motions of the $K$ agents 
along a circular chain of environment pseudo-spins: Controlled by position a 
local transformation $\hat{U}_{\alpha_{k}}^{(Sk)}$ is followed by a 
pair-interaction $\hat{U}_{\theta}^{(Sk,\mu)},\;\theta=0,\pi$, and vice versa. 
For any given agent $Sk$, 
the subscript $\theta$~(``type of the agent'') is assumed to be 
fixed. The various agents may move in a constant or in a 
changing sequential order. In the former case a strict iterative map results 
from $2M$~unitary transformations for each agent (equivalent to one full cycle 
$p=1,2, \cdots$). For $K=1$ this architecture may be viewed as a 
quantum-Turing-machine~\cite{DEU85,DEU89,BEN82,BEN96,BEN98}, 
with the agent-spin being the Turing-head and the environment acting as the 
Turing-tape. Most results will be presented for this simplest scenario. 
In this case the resulting iterative map can be specified as
\ba
|\psi(m_1)\rangle &=& \hat{U}^{(S1)}(n_{1}) \cdots \hat{U}^{(S1)}(1) 
\bigl(\hat{U}^{(S1)}(2M) \cdots \hat{U}^{(S1)}(1)\bigr)^{p_{1}} |\psi(0)\rangle
\nonumber\\
&\equiv& \hat{T}^{(S1)}(m_{1})\,|\psi(0)\rangle \label{psi_m}
\ea
where $n_{1}=1,2, \cdots, 2M;\,m_{1}=n_{1}+2M(p_{1}-1);\,p_{1}$ being the 
number of completed cycles, $m_{1}$ the step number, and 
\ba
\hat{U}^{(S1)}(2\mu-1) &:=& \hat{U}_{\alpha_{\mu}}^{(S1)}\nonumber\\
\hat{U}^{(S1)}(2\mu) &:=& \hat{U}_{\theta}^{(S1,\mu)}\;. \label{U_m}
\ea

\section*{Pure-state trajectories for $K=1, \theta=0$}
We restrict ourselves to the reduced state-space dynamics 
(our ``system of interest'')
\be
\lambda_{i}^{(S1)}(m_{1}) = Q_{i00 \cdots 0}(m_{1}) = \langle\psi(m_{1})|
\hat{Q}_{i00 \cdots 0}|\psi(m_{1})\rangle\;.
\ee
Due to entanglement we will, in general, see the apparent decoherence, 
\be
Y_{1}^{(S1)} = \sum_{i=1}^{3} |\lambda_{i}^{(S1)}|^{2} < 1\;.
\ee
However, for specific initial states~$|\psi(0)\rangle$ the state of the 
Turing-head~$S1$ will remain pure: As $|\pm\rangle\!^{(\mu)} = 
\frac{1}{\sqrt{2}}\bigl(|0\rangle\!^{(\mu)} \pm 
|1\rangle\!^{(\mu)}\bigr)$ are the eigenstates of 
$\hat{\lambda}_{1}^{(\mu)}$ with 
$\hat{\lambda}_{1}^{(\mu)} |\pm\rangle\!^{(\mu)} = \pm 
|\pm\rangle\!^{(\mu)}$, the action of the QCNOT reduces to
\ba
\hat{U}_{0}^{(S1,\mu)} |\varphi\rangle\!^{(S1)} \otimes\,|+\rangle\!^{(\mu)} 
&=& |\varphi\rangle\!^{(S1)} \otimes\,|+\rangle\!^{(\mu)}\nonumber\\
\hat{U}_{0}^{(S1,\mu)} |\varphi\rangle\!^{(S1)} \otimes\,|-\rangle\!^{(\mu)} 
&=& \hat{\lambda}_{3}^{(S1)} |\varphi\rangle\!^{(S1)} \otimes\,
|-\rangle\!^{(\mu)}\;.
\ea
As a consequence, for the initial product-state
\be
|\psi(0)\rangle =  |\varphi(0)\rangle\!^{(S1)} \otimes\,
|{\mathcal{P}}^{j}(0)\rangle
\ee
with
\ba
&|\varphi(0)\rangle\!^{(S1)} = \cos{(\varphi_{0}/2)} |0\rangle\!^{(S1)} - 
i \sin{(\varphi_{0}/2)} |1\rangle\!^{(S1)}&\nonumber\\ 
&|{\mathcal{P}}^{j}(0)\rangle \in \bigl\{|\pm\rangle\!^{(1)} \otimes\,
|\pm\rangle\!^{(2)} \otimes \cdots \otimes\,|\pm\rangle\!^{(M)}\bigr\}\;,&
\ea
the state~$|\psi(m)\rangle$ remains a product-state at all steps~$m$ 
(``primitives''):
\be
|\psi(m|{\mathcal{P}}^{j})\rangle = |\varphi(m|{\mathcal{P}}^{j})
\rangle\!^{(S1)} \otimes\,|{\mathcal{P}}^{j}(0)\rangle\;.
\ee
Here $|\varphi(m|{\mathcal{P}}^{j})\rangle\!^{(S1)}$ is the Turing-head state 
at step~$m$ conditioned by $|{\mathcal{P}}^{j}(0)\rangle$ and for a given 
head-state there are $2^M$~primitives. The Turing-head Bloch-vectors 
\be
\lambda_{k}^{(S1)}(m|{\mathcal{P}}^{j}) = \langle\psi(m|{\mathcal{P}}^{j})| 
\hat{\lambda}_{k}^{(S1)} \otimes \hat{1}^{(1)} \otimes \hat{1}^{(2)} \cdots 
\otimes \hat{1}^{(M)} |\psi(m|{\mathcal{P}}^{j})\rangle
\ee
are all confined to the\, {$k=2,3$-plane} and obey the relation 
$|\vec{\lambda}^{(S1)}(m|{\mathcal{P}}^{j})| = 1$. In Fig~\ref{fig_primitive} 
we show the 
$4$~primitives $\lambda_{k}^{(S1)}(m|{\mathcal{P}}^{j}),\;j=1,2,3,4$, for 
$M=2$ and $\alpha=\pi/\sqrt{3}$. 
The primitives are either periodic (Floquet-states) and then independent of 
$\alpha$, or aperiodic (here: 
${\mathcal{P}}^{++}$) and then controlled by $\alpha$ (the latter orbits are 
also periodic if $\alpha$ is a rational multiple of $\pi$). A similar 
analysis holds for $\theta=\pi$.

\section*{Superposition of primitives for $K=1, \theta=0$}
Now let the initial state be
\be
|\psi(0)\rangle = \sum_{j=1}^{2^M} a_{j} |\varphi(0)\rangle\!^{(S1)} \otimes\,
|{\mathcal{P}}^{j}(0)\rangle\;.
\ee
Then we find at step~$m$
\be
|\psi(m)\rangle = \sum_{j=1}^{2^M} a_{j} |\varphi(m|{\mathcal{P}}^{j})
\rangle\!^{(S1)} \otimes\,|{\mathcal{P}}^{j}(0)\rangle
\ee
and, observing the orthogonality of the $|{\mathcal{P}}^{j}(0)\rangle$,
\ba
\lambda_{k}^{(S1)}(m) &=& \langle\psi(m)| \hat{\lambda}_{k}^{(S1)} \otimes 
\hat{1}^{(1)} \otimes \hat{1}^{(2)} \otimes \cdots \otimes \hat{1}^{(M)} 
|\psi(m)\rangle\nonumber\\
&=& \sum_{j=1}^{2^M} |a_{j}|^{2}\,\lambda_{k}^{(S1)} (m|{\mathcal{P}}^{j})\,.
\label{lambda_S1}
\ea
This trajectory of agent $S1$ represents a non-orthogonal pure-state 
decomposition with weights $|a_{j}|^{2}$ independent of $m$, and the 
decomposition can be seen as an intuitive example for 
quantum parallelism: The individual 
Turing-head performs exponentially many primitive trajectories 
``in parallel''. Different initial states with the same $|a_{j}|^2$ 
show the same reduced dynamics for $S1$.
 
Special superpositions are 
\begin{eqnarray}
&|a_j|^2\,= & \left\{
 \begin{array}{cl}
 \mbox{const.} &\;\; \mbox{for}\; j \in \mbox{\{periodic orbits\}}\\
 0 &\;\; \mbox{otherwise}
\end{array}
\right.
\end{eqnarray}
and the complementary type
\begin{eqnarray}
&|a_j|^2\,= & \left\{
 \begin{array}{cl}
 \mbox{const.} &\;\; \mbox{for}\; j \in \mbox{\{aperiodic orbits\}}\\
 0 &\;\; \mbox{otherwise}\;.
\end{array}
\right.
\end{eqnarray}
The example for $M=3$ is shown in Fig~\ref{mmm}. 
In this case each superposition 
consists of $4$ primitives. The trend towards reduced Bloch-vector lengths is 
easily recognized; however, there are no privileged basis states (in which 
case the points would be on a single straight line). 

Finally, starting from the ground state, the typical initial state for 
quantum computation~\cite{DIV95,EKE96}
\be
|\psi(0)\rangle = |0\rangle\!^{(S1)} \otimes\,|00 \cdots 0\rangle
\ee
all $2^M$ pure state trajectories contribute with equal weight
\be
|a_j|^2 = \frac{1}{2^M}\;.\nonumber
\ee
Calculations for $M=3,10$ are shown in Fig~\ref{spindown}. 
The resulting pattern 
(``quasi-$1$-dimensional point manifolds'') shows the existence of a local 
invariant (for $M=1,2$, {\it{cf.}}~\cite{KIM99}), 
which is a consequence of the 
underlying primitives.

\section*{Computational reducibility}
Knowing the $2^M$ primitives we can calculate the reduced Turing-head 
dynamics from eq.~(\ref{lambda_S1}) for any initial tape-state. 
This procedure, to be sure, 
becomes prohibitive for large $M$, as the number of those primitives grows 
exponentially. It turns out that a complementary problem is much easier to 
solve: To calculate all possible iterative maps (for any $M$ and any 
control-angles $\alpha$) for a {\em{selected}} initial tape-state 
(here: the ground state). 
This can be done by the recursion relation
\begin{eqnarray}
\lambda_{1}^{(S1)}(m) &=& 0 \nonumber \\ 
\lambda_{2}^{(S1)}(m) &=& Y_{m,M} \nonumber\\
\lambda_{3}^{(S1)}(m) &=& Z_{m,M}
\end{eqnarray}
where $Y_{m,M}$ and $Z_{m,M}$ are specified in Table~\ref{recursion}.

\section*{Several agents, $K \geq 2$}
We start by noting the commutator relations: 
\begin{eqnarray}
\bigl\lbrack\hat{U}_{\theta}^{(Sk,\mu)}, 
\hat{U}_{\theta}^{(Sk',\mu')}\bigr\rbrack &=& 
0\nonumber\\
\bigl\lbrack\hat{U}_{0}^{(Sk,\mu)}, \hat{U}_{\pi}^{(Sk',\mu')}\bigr\rbrack &=& 
\left\{
 \begin{array}{ll}
 -2 \hat{P}_{00}^{(Sk)} \otimes \hat{\lambda}_{3}^{(\mu)} \delta_{\mu\mu'} 
&\;\; \mbox{for}\; Sk=Sk'\\
 -2 \hat{P}_{00}^{(Sk)} \otimes \hat{P}_{00}^{(Sk')} \otimes 
\hat{\lambda}_{3}^{(\mu)} \delta_{\mu\mu'} &\;\; \mbox{for}\; Sk\ne Sk'\;.
\end{array}
\right.
\end{eqnarray}
This means that for any agent of the same type {\em{all}} unitary 
transformations 
between different agents $Sk \ne Sk'$ commute, i.e. the (time-) ordering is 
irrelevant! Thus for $K=2$, e.g., one finds ({\it{cf.}} eq.~(\ref{psi_m}))
\ba
|\psi(m_1,m_2)\rangle &=& \cdots\,\hat{U}^{(S2)}(4)\,\hat{U}^{(S2)}(3)\,
\hat{U}^{(S1)}(4)\,\hat{U}^{(S1)}(3)\,\times\nonumber\\
&& \hat{U}^{(S2)}(2)\,\hat{U}^{(S2)}(1)\,
\hat{U}^{(S1)}(2)\,\hat{U}^{(S1)}(1)\,|\psi_0\rangle\nonumber\\
&=& \underbrace{\cdots\,\hat{U}^{(S2)}(4)\,\hat{U}^{(S2)}(3)\,
\hat{U}^{(S2)}(2)\,\hat{U}^{(S2)}(1)}_{\hat{T}^{(S2)}(m_2)}\,\times\nonumber\\
&& \underbrace{\cdots\,\hat{U}^{(S1)}(4)\,\hat{U}^{(S1)}(3)\,
\hat{U}^{(S1)}(2)\,\hat{U}^{(S1)}(1)}_{\hat{T}^{(S1)}(m_1)}\,|\psi_0\rangle
\nonumber\\
&=& \hat{T}^{(S1)}(m_1)\,\hat{T}^{(S2)}(m_2)\,|\psi(0)\rangle
\ea
where $\hat{U}^{(S2)}(2\mu-1), \hat{U}^{(S2)}(2\mu)$ are equivalent to 
$\hat{U}^{(S1)}(2\mu-1), \hat{U}^{(S1)}(2\mu)$ (eq.~(\ref{U_m})) respectively, 
so that the local agent properties are independent of each other:
\ba
\lambda_{j}^{(S1)}(m_1) &=& \langle\psi(0)\hat{T}^{\dagger(S1)}(m_1) 
\hat{T}^{\dagger(S2)}(m_2)|\hat{\lambda}_{j}^{(S1)}|\hat{T}^{(S2)}(m_2) 
\hat{T}^{(S1)}(m_1)\psi(0)\rangle\nonumber\\
&=& \langle\psi(0)\hat{T}^{\dagger(S1)}(m_1)|\hat{\lambda}_{j}^{(S1)}|
\hat{T}^{(S1)}(m_1)\psi(0)\rangle\;.
\ea
Corresponding relations hold for $S2$. As a consequence the patterns of each 
agent are not influenced by the presence of the other, even though both agents 
become entangled, in addition to the entanglement between each agent and the 
environment (tape)! 

Things change dramatically if the two agents are of 
different type. Then the order of the unitary transformations with 
respect to each agent matters, and the actions for each agent can no longer 
be grouped together. In Fig~\ref{two_agents} we show the result for 
$K=2, M=2$. Now the 
patterns for each agent show an (apparent) randomness, there are no longer 
obvious local invariants visible.

\section*{Summary and conclusions}
Open systems are well-known examples exhibiting loss of coherence with 
respect to certain privileged basis states (the ``measurement basis''). 
Here we have considered an all-quantum-mechanical spin-network subject to a 
discretized unitary transformation (iterative map). We deliberately choose to 
``look'' only at a subset of spins, the $K$ ``agents'', which interact with a 
closed chain of $M$ ``environment''-spins, one after the other. For this 
pure-state dynamics the ``flow of coherence'' is explicitly demonstrated as 
well as the entanglement-induced decoherence. For agents of the same type 
fascinating patterns result in their respective Bloch-spheres (a quantum 
version of Poincar\'{e}-cuts). These patterns can be understood as being based 
on a set of pure-state trajectories (``primitives''). For specific initial 
states even explicit recursion relations (in the agent Bloch-vector-space) 
exist. This remarkable computational reducibility works for any network-size 
$N=K+M$. However, if the agents are of different type (non-commuting), they 
disturb each other via their common environment, and the resulting 
Bloch-sphere patterns show apparent randomness without obvious signs for 
invariants. This indicates that geometrical aspects can be a useful 
supplement to conventional algebraic results for the discussion of 
decoherence and entanglement.

\section*{Acknowledgments}
We would like to thank C. Granzow, A. Otte and R. Wawer for stimulating 
discussions. 

\section*{Figure captions}
\begin{figure} 
\refstepcounter{figure}\label{architecture}
\vspace*{14.7cm}
\begin{center}
\includegraphics{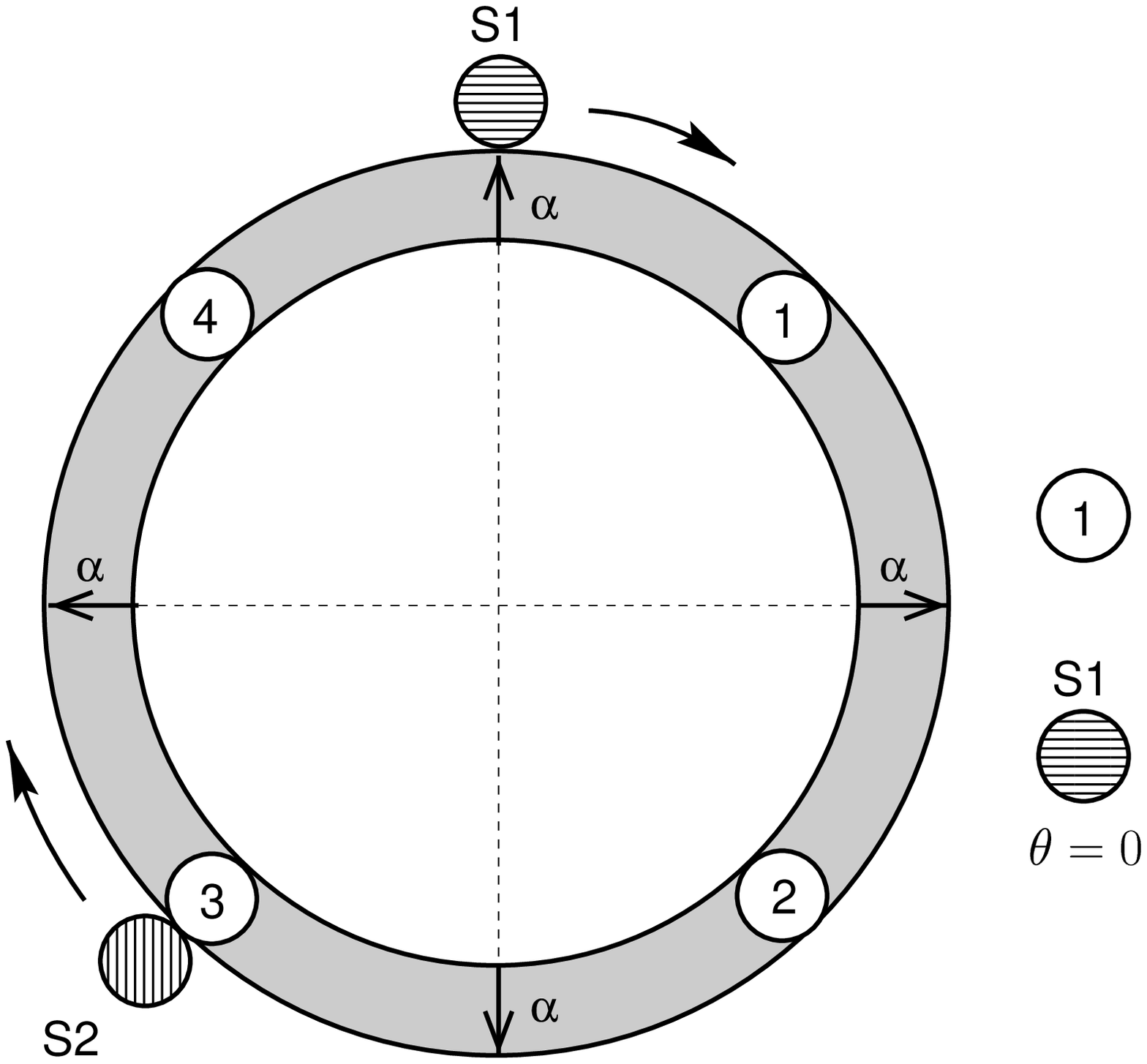}
\end{center}
\vspace*{-4.5cm}
\end{figure}

Fig.~\ref{architecture}\, The network architecture: 
The moving agents $S1, S2$ move along the 
circular environment (spins $1,2,3,4$) in discrete steps, thus iterating 
between local transformations (arrows) and pair interactions (when in touch 
with an environment spin).\\

\begin{figure} 
\refstepcounter{figure}\label{fig_primitive}
\vspace*{12.7cm}
\begin{center}
\includegraphics{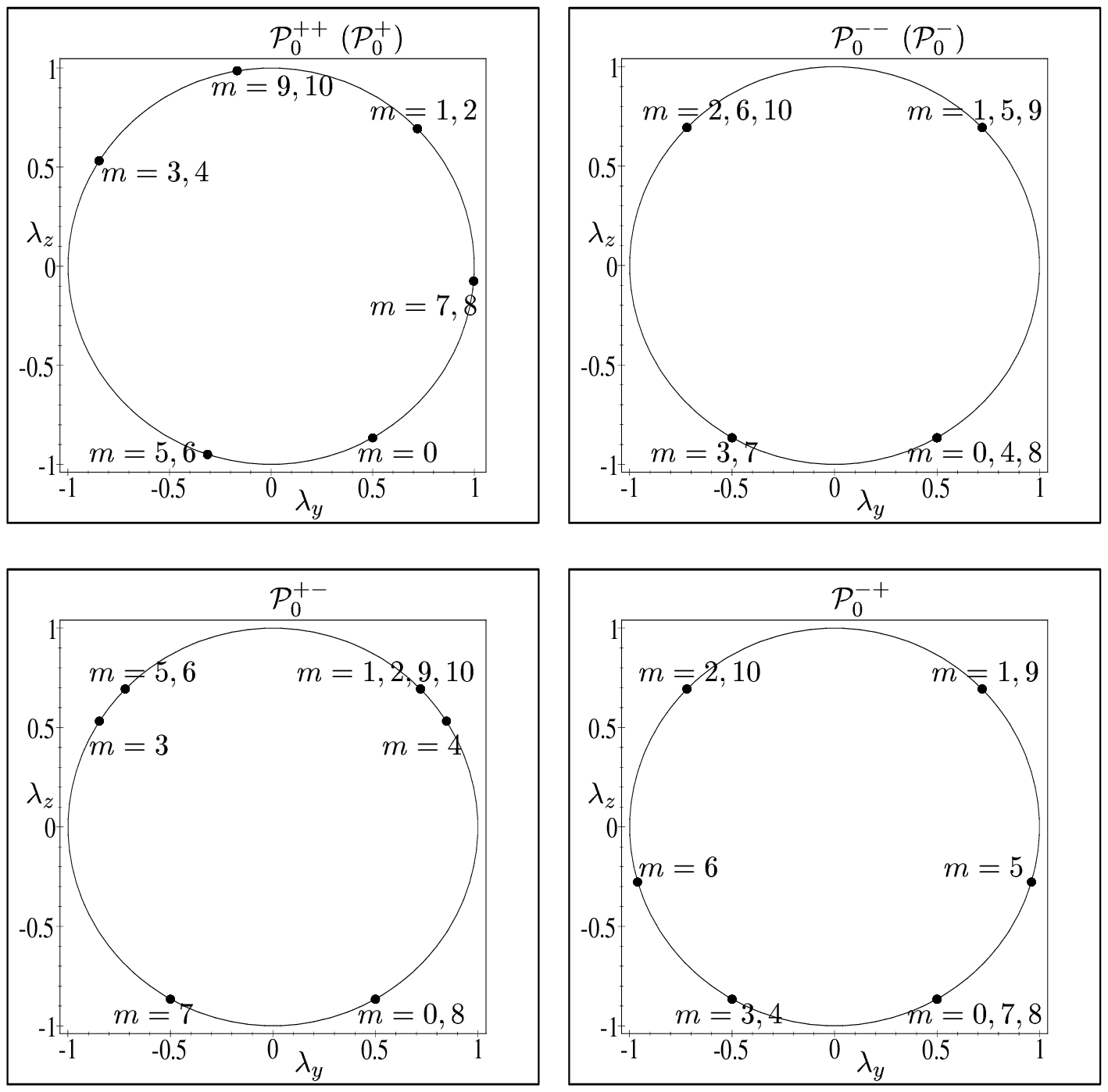}
\end{center}
\vspace*{-5.5cm}
\end{figure}

Fig.~\ref{fig_primitive}\, The primitives ${\mathcal{P}}_{0}^{+}$ (aperiodic) 
and ${\mathcal{P}}_{0}^{-}$ (periodic) for\, $K=1, \theta=0, M=1$, and\, 
${\mathcal{P}}_{0}^{++}$ (aperiodic), 
${\mathcal{P}}_{0}^{--}$, ${\mathcal{P}}_{0}^{+-}$, ${\mathcal{P}}_{0}^{-+}$ 
(periodic) for $K=1, l=1, M=2$; {$\alpha=\pi/\sqrt{3}$},\, 
{$\varphi(0)=\pi/6$}.\\

\begin{figure} 
\refstepcounter{figure}\label{mmm}
\vspace*{13.1cm}
\begin{center}
\includegraphics{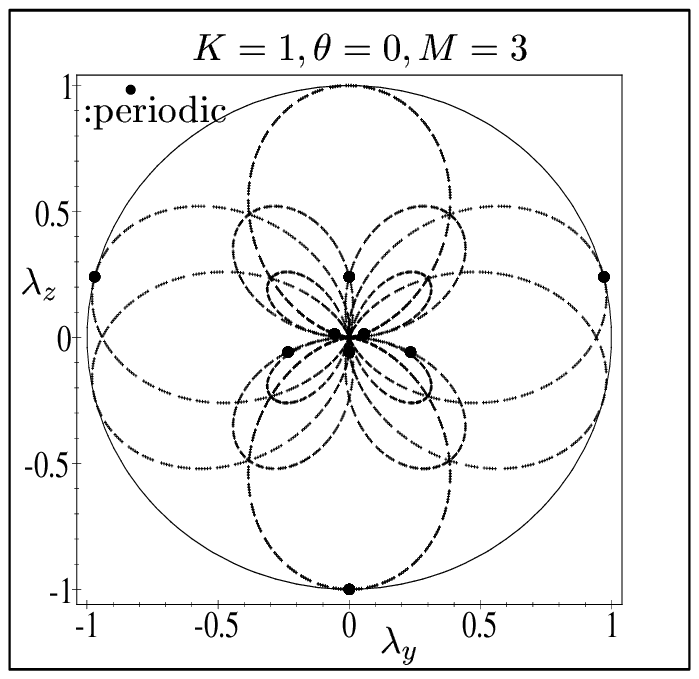}
\end{center}
\vspace*{-10cm}
\end{figure}

\begin{figure} 
\refstepcounter{figure}\label{spindown}
\vspace*{12.3cm}
\begin{center}
\includegraphics{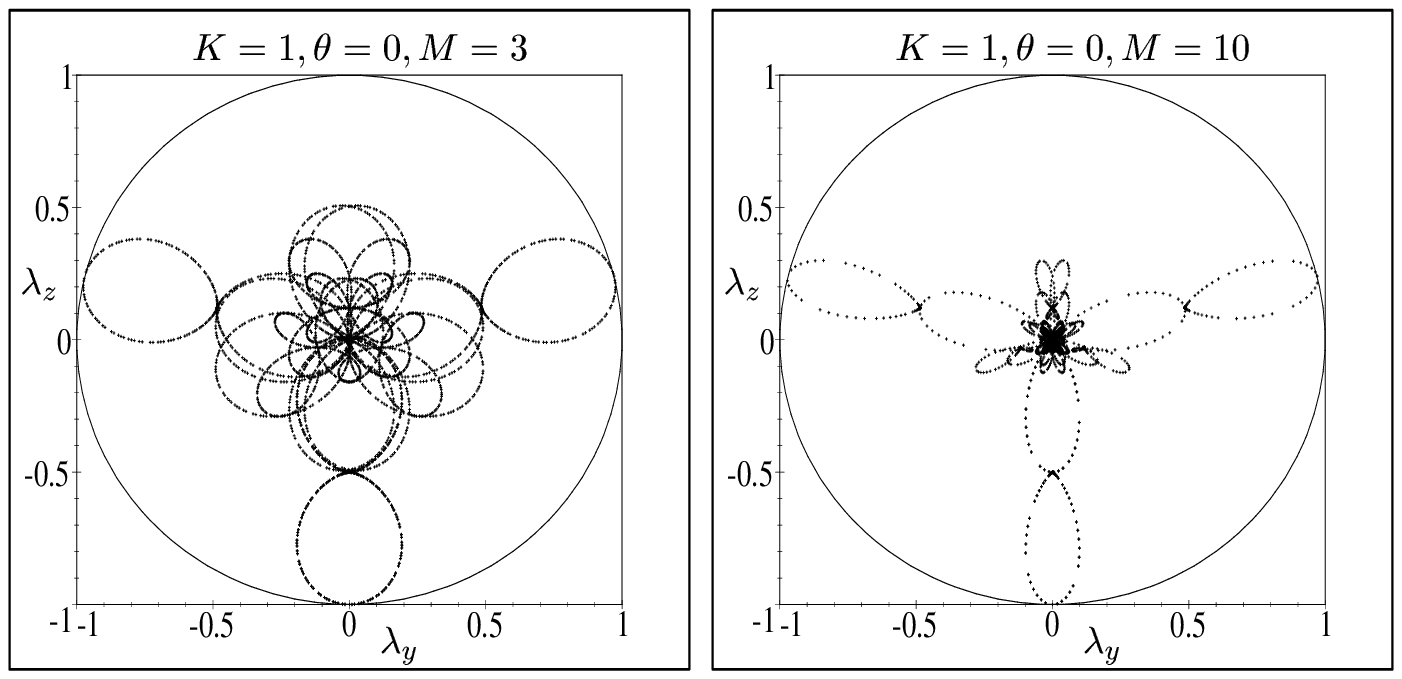}
\end{center}
\vspace*{-9cm}
\end{figure}

Fig.~\ref{mmm}\, Equal-weight superpositions ($a_{j}=1/2$) of $4$~periodic 
($4$~aperiodic) orbits for $|\psi_{0}\rangle=|0\rangle\!^{(S1)}\otimes\,
|000\rangle, K=1, \theta=0, M=3$, 
and step numbers $m\leq 3000$. The equal-weight superposition ($1/\sqrt{2}$) 
of these two, in turn, generates the pattern for 
$|\psi_{0}\rangle=|0\rangle\!^{(S1)}\otimes\,|000\rangle$ 
(see Fig~\ref{spindown} for $M=3$); $\alpha=\pi/\sqrt{3}$.\\

Fig.~\ref{spindown}\, Turing-head-patterns for 
$|\psi_{0}\rangle=|0\rangle\!^{(S1)}
\otimes\,|00 \cdots 0\rangle$, $K=1, l=1, M=3,10$, and step numbers 
$m\leq 3000$; $\alpha=\pi/\sqrt{3}$.\\

Fig.~\ref{two_agents}\, Local Bloch-vector patterns, 
$\lambda_y=\lambda_{y}^{(S1)}, \lambda_z=\lambda_{z}^{(S1)}$. 
First row: $K=1, M=1$; $\theta=0$ for $S1$ (left),\, $\theta=\pi$ (right); 
second row: $K=2, \theta=0$~for~$S1$,\, $\theta=\pi$~for~$S2$; 
$M=1$ (left), $M=2$ (right); last row: $K=1, M=2$; $\theta=0$~for~$S1$ 
(left),\, $\theta=\pi$ (right); 
$|\psi_{0}\rangle=|00 \cdots 0\rangle$; $\alpha=\pi/\sqrt{3}$ in each case, 
$m_i\leq 4500,\,i=S1,S2$.


\begin{table} 
\caption{Recursion relations for the reduced state evolution of $S1$ in 
the case of $K=1$, $\theta=0,\,|\psi_{0}\rangle=|0\rangle\!^{(S1)}\otimes\,
|00 \cdots 0\rangle$. Let $Y_{m}=Y_{m,M}$, 
$Z_{m}=Z_{m,M}$, $Z_{m,0}:=-1$, and $m':=m -4p + 2$, where $p$ is the 
cycle number for step $m$; $m=n+2M(p-1)$, $n=1,2, \cdots, 2M$. $Y_{0}=0$, 
$Y_{1}=\sin{\alpha}$, $Z_{0}=-1$, $Z_{1}=-\cos{\alpha}$. \label{recursion}} 
\begin{center}
\renewcommand{\arraystretch}{1.4}
\setlength\tabcolsep{25pt}
\begin{tabular}{c|l}
\hline
$Y_{m}=-Y_{1}Z_{m-1}-Z_{1}Y_{m-1}$ & $n=$ odd\\
\hline
$Y_{m,M}=Y_{m-1,M}+Y_{1}Z_{m',M-2}$ & $n=$ even $\ne 2M$\\
\hline
$Y_{m,M}=Y_{m-1,M}-Y_{1}(-Z_{1})^{M-1}$ & $n=2M$, $p=$ odd\\
\hline
$Y_{m,M}=Y_{m-1,M}$ & $n=2M$, $p=$ even\\
\hline
$Z_{m} = -Z_{1} Z_{m-1}+ Y_{1} Y_{m-1}$ & $n=$ odd\\
\hline
$Z_{m} = -Z_{1} Z_{m-2}+ Y_{1} Y_{m-2}$ & $n=$ even\\
\hline
\end{tabular}
\end{center}
\end{table}
\newpage
\begin{figure} 
\refstepcounter{figure}\label{two_agents}
\vspace*{10cm}
\begin{center}
\includegraphics{figure5.ps}
\end{center}
\vspace*{.7cm}
\end{figure}
\end{document}